\documentclass[a4paper,fleqn]{cas-sc}

\usepackage[numbers]{natbib}
\usepackage[most]{tcolorbox}
\usepackage{tikz}
\usetikzlibrary{positioning}
\usepackage{amsmath, amssymb}
\def\tsc#1{\csdef{#1}{\textsc{\lowercase{#1}}\xspace}}
\tsc{WGM}
\tsc{QE}
\tsc{EP}
\tsc{PMS}
\tsc{BEC}
\tsc{DE}

\begin{document}
\let\WriteBookmarks\relax
\def\floatpagepagefraction{1}
\def\textpagefraction{.001}
\shorttitle{OTU Separability in Multinomial CoDA}
\shortauthors{R. Alberich et~al.}

\title [mode = title]{Identification of Separable OTUs for  Multinomial Classification in Compositional Data Analysis}  
%
\tnotemark[1]

\tnotetext[1]{R. Alberich, I. García, A. Mir and F. Rosselló acknowledge support from the research project PID2021-126114NB-C44
funded by MCIN/AEI/ 10.13039/501100011033 and by “ERDF A way of making Europe”.}


\author[1,2]{R. Alberich}
\ead{r.alberich@uib.es}

\author[1,2]{N. A. Cruz}
\ead{nelson-alirio.cruz@uib.es}

\author[1]{R. Fernández}
\ead{r.fernandez@uib.es}

\author[1,2]{I. García}
\ead{irene.garcia@uib.es}

\author[1,2]{A. Mir}
\ead{arnau.mir@uib.es}


\author[1]{F. Rosselló}
\ead{cesc.rossello@uib.es}

\affiliation[1]{organization={Department of Math and Computer Science (University of Balearic Islands), IDISBA Research Institute},
                addressline={Crta. Valldemossa, km. 7,5}, 
                city={Palma},
                postcode={07122}, 
                country={Spain}}

\affiliation[2]{organization={Department of Math and Computer Science, Artificial Intelligence Research Institute (University of Balearic Islands), IDISBA Research Institute, },
                addressline={Crta. Valldemossa, km. 7,5}, 
                city={Palma},
                postcode={07122}, 
                country={Spain}}
\cortext[cor1]{Corresponding author}


\begin{abstract}
High-throughput sequencing has transformed microbiome research, but it also produces inherently compositional data that challenge standard statistical and machine learning methods. Identifying microbial taxa that discriminate between biological or clinical groups requires methods that both respect compositional constraints and provide rigorous statistical inference. In this work, we propose a multinomial classification framework for compositional microbiome data based on penalized log-ratio regression and pairwise separability screening. The method quantifies the discriminative ability of each OTU through the area under the receiver operating characteristic curve ($AUC$) for all pairwise log-ratios and aggregates these values into a global separability index $S_k$, yielding interpretable rankings of taxa together with confidence intervals.

We illustrate the approach by reanalyzing the Baxter colorectal adenoma dataset and comparing our results with Greenacre’s ordination-based analysis using Correspondence Analysis and Canonical Correspondence Analysis. Our models consistently recover a core subset of taxa previously identified as discriminant, thereby corroborating Greenacre’s main findings, while also revealing additional OTUs that become important once demographic covariates are taken into account. In particular, adjustment for age, gender, and diabetes medication improves the precision of the separation index and highlights new, potentially relevant taxa, suggesting that part of the original signal may have been influenced by confounding.

Overall, the integration of log-ratio modeling, covariate adjustment, and uncertainty estimation provides a robust and interpretable framework for OTU selection in compositional microbiome data. The proposed method complements existing ordination-based approaches by adding a probabilistic and inferential perspective, strengthening the identification of biologically meaningful microbial signatures.
\end{abstract}


\maketitle

\section{Introduction}

High-throughput sequencing technologies have made it possible to characterize microbial communities in great detail, typically through counts of operational taxonomic units (OTUs) (Caporaso et al., 2011). The resulting data are inherently compositional: each sample is constrained to a fixed total, a property known as the “closure” constraint (Aitchison, 1986; Gloor et al., 2017). As a consequence, microbiome datasets generated by high-throughput sequencing should be treated as compositional at every stage of the analysis (Gloor et al., 2017; Fernandes et al., 2014). When standard multivariate methods are applied directly to raw or proportion-normalized counts, they can introduce spurious associations and lead to misleading inferences (Gloor et al., 2017). This issue is particularly problematic in supervised learning, where the goal is often to identify a subset of OTUs that best discriminates between multiple phenotypic or clinical groups.

The challenge of identifying discriminative features from compositional data is not unique to microbiome research. Compositional structures arise in many other fields, including geochemistry, metabolomics, ecology, single-cell transcriptomics, and economics, whenever measurements represent parts of a whole (Greenacre, Martínez-Álvaro \& Blasco, 2021). In all these settings, statistical and machine learning methods that ignore the compositional constraint risk producing biased estimates and incorrect interpretations (Quinn et al., 2018). There is therefore a clear need for classification and feature-selection frameworks that explicitly account for compositionality, in order to support valid inference, interpretability, and reproducibility in high-dimensional applications.

In microbiome studies in particular, researchers are often interested not only in accurate prediction but also in interpretable variable selection—identifying “separable” OTUs whose relative abundances differ systematically across groups and that can be combined into discriminative microbial signatures (Susin et al., 2020). Existing approaches typically rely on marginal testing, methods developed for binary outcomes, or procedures that do not fully respect the compositional structure of the data (Susin et al., 2020). Moreover, they rarely offer a coherent way to quantify uncertainty in the selected OTUs or to propagate OTU-level variability into global measures of separability in multinomial classification problems.

In this work, we introduce a classification framework specifically designed for compositional microbiome data, with the aim of identifying microbial taxa that most effectively discriminate among multiple biological or clinical groups. The framework starts with standard preprocessing steps to ensure data quality and comparability, including the removal of rare taxa, Bayesian-multiplicative imputation of zeros, and centered log-ratio transformation to properly handle compositional constraints (Martín-Fernández et al., 2015; Gloor et al., 2021).

At its core, the methodology uses a regression-based model that relates microbial composition to group membership while accounting for the inherent interdependence between components. Extending this model to the multinomial setting allows us to work naturally with more than two groups. By expressing the model in terms of log-ratios between pairs of OTUs, we can directly examine how the balance between taxa contributes to group differentiation (Greenacre et al., 2021; Hinton et al., 2022).

Building on this formulation, we propose a pairwise log-ratio screening procedure that measures the discriminative power of each OTU pair. For every pair, we assess its ability to separate groups using the area under the receiver operating characteristic curve (AUC) as a measure of separability. These pairwise results are then aggregated into a global separability index that summarizes the overall contribution of each OTU across all pairs, providing a data-driven ranking of the most informative taxa.

Finally, we incorporate formal procedures to quantify the uncertainty associated with separability estimates, combining analytical variance estimators with resampling-based inference. This allows us to ensure that the OTUs we identify are not only discriminative but also statistically robust. The complete framework is implemented in the \texttt{codabiocom} R package \cite{CALLE2023}, which offers tools for covariate adjustment, uncertainty quantification, and parallel computation, which makes it suitable for large-scale compositional datasets.

\section{Preliminaries}

The microbiome datasets obtained through high-throughput sequencing consist of abundance counts for operational taxonomic units (OTUs) $m$ in $n$ samples. These data are compositional by nature: only relative abundances are meaningful, and the total abundance within each sample is constrained to a fixed sum. Throughout this section, we describe the preprocessing steps and fundamental definitions used in our framework for OTU classification and selection.

\subsection{Data Preprocessing}

To ensure robustness and comparability between samples, the abundance matrix $\mathbf{X}$ of dimension $n \times m$ undergoes a series of preprocessing steps:

\begin{description}
    \item[\textbf{Filtering of rare OTUs.}] OTUs with extremely low prevalence contribute little information and increase noise. Those with two or fewer non-zero counts are excluded from the analysis, i.e., OTU $j$ is removed if $\sum_{l=1}^n x_{lj} \le 2$.

    \item[\textbf{Zero imputation.}] Because compositional data analysis (CoDA) relies on logarithmic transformations, zero counts must be replaced with small positive values. We employ the Bayesian-multiplicative imputation proposed by \cite{BAYZEROSIMP}, which preserves the sample composition while maintaining relative ratios among non-zero components.

    \item[\textbf{Centered log-ratio (clr) transformation.}]  
    After imputation, each sample $\mathbf{x}_{i\cdot}$ is transformed using the clr transformation:
    $$
        \tilde{x}_{ij} = \log\!\left(\frac{x_{ij}}{\big(\prod_{l=1}^m x_{il}\big)^{1/m}}\right),
    \quad
        i = 1,\dots,n,\; j = 1,\dots,m.
    $$
    This yields the transformed matrix $\mathbf{\tilde{X}}$, ensuring scale invariance and compatibility with Euclidean geometry.
\end{description}

The first step of the workflow (Step~1 in Figure~\ref{fig:overview}) 
involves removing low-frequency OTUs, imputing zeros and applying the 
centered log-ratio (clr) transformation.

\subsection{Concept of separable OTUs}

In this paper we use the term separable OTUs to denote taxa that, taken together, allow for a good discrimination between the outcome classes in a compositional setting. Let \(\mathcal{G} = \{1,\ldots,G\}\) denote the set of classes and let \(\mathcal{J} = \{1,\ldots,m\}\) index the OTUs. For any pair of OTUs \(j,j' \in \mathcal{J}\), we consider the log-ratio
\[
z_{ijj'} = \log\frac{x_{ij}}{x_{ij'}},
\]
where \(\mathbf{x}_i = (x_{i1},\ldots,x_{im})\) denotes the composition of sample \(i\). A pair \((j,j')\) is said to be *separable* for a given classification task if the distribution of \(z_{ijj'}\) differs substantially across outcome classes.

To quantify separability we fit, for each ordered pair \((j,j')\), a (multinomial) logistic regression model with \(z_{ijj'}\) as predictor and the outcome as response, and compute the corresponding generalized AUC. This yields a matrix of pairwise AUC values, which we denote by
\[
A = \bigl(\text{AUC}_{jj'}\bigr)_{j,j'=1}^m.
\]
High values of \(\text{AUC}_{jj'}\) indicate that the log-ratio between OTUs \(j\) and \(j'\) provides good discrimination between classes; low values indicate limited discriminatory power.

\section{Separability index for a set of OTUs}

Our goal is to rank OTUs not only individually, but also in terms of how well *sets* of OTUs perform in combination. For any subset \(S \subseteq \mathcal{J}\) with \(|S| = k\), we define its *separability index* as
\[
\mathrm{Sep}(S) = \frac{1}{k(k-1)} \sum_{j \in S} \sum_{\substack{j' \in S \\ j' \neq j}} \text{AUC}_{jj'},
\]
that is, the average of all pairwise AUC values associated with the OTUs in \(S\). Intuitively, a set \(S\) will have a high separability index if most of its log-ratio pairs induce good discrimination between the outcome classes.

In practice, we construct a ranking of OTUs, \(j_{(1)}, j_{(2)}, \ldots, j_{(m)}\), where \(j_{(1)}\) denotes the most individually discriminant OTU, \(j_{(2)}\) the second most discriminant, and so on (see Section X.X for details). For each \(k = 2,\ldots,k_{\max}\) we then consider the top-\(k\) set
\[
S_k = \{j_{(1)},\ldots,j_{(k)}\}
\]
and compute its separability index
\[
S_k = \mathrm{Sep}(S_k).
\]
The curve \(k \mapsto S_k\) describes how average pairwise separability improves (or deteriorates) as more OTUs are included. We choose the number of OTUs \(\hat{k}\) as the value maximizing \(S_k\),
\[
\hat{k} = \arg\max_k S_k,
\]
and refer to the OTUs in \(S_{\hat{k}}\) as the *separable OTUs* for the classification task under study.

This definition is fully determined by the AUC matrix \(A\) and does not rely on any particular clustering or dendrogram representation. We therefore focus on \(S_k\) as our main separability index throughout the rest of the paper.

\section{Binary classification framework}

We first describe our methodology in the binary setting, which forms the basis for its extension to multinomial classification. Let $\mathbf{X}$ be the $n \times m$ abundance matrix, with rows corresponding to samples and columns corresponding to OTUs, and let $Y$ denote the group indicator taking values in $\{0,1\}$. Our goal is to identify those OTUs (or combinations thereof) that best discriminate between the two groups, while respecting the compositional nature of the data.

\subsection{CoDA-based regression for microbiome analysis}
\label{subsec:coda_regression}

We address the classification problem using a penalized regression model in which discriminative OTUs are those with nonzero regression coefficients. More specifically, we build on the microbiome signature based on log-ratio analysis proposed in \cite{CALLE2023}.

We introduce the binary response variable $Y$ as

$$
Y=
\begin{cases}
    1, & \text{if sample } \mathbf{x}_{i\cdot} \in g_1,\\
    0, & \text{if sample } \mathbf{x}_{i\cdot} \in g_2,
\end{cases}
$$

where each sample $\mathbf{x}_{i\cdot}$ corresponds to the $i$-th row of matrix $\mathbf{X}$. The objective is to model $Y$ based on the information in $\mathbf{X}$ within a generalized linear model framework that incorporates the compositional structure of the data.

A natural starting point is the log-contrast model \citep{AITCHISON1984,LU2019}, defined as
$$
g(E(Y)) = \alpha_0 + \sum_{j=1}^m \alpha_j \log(X_j),
$$

subject to the zero-sum constraint $\displaystyle\sum_{j=1}^m \alpha_j = 0$.

This constraint enforces the scale-invariance principle required in compositional data analysis (CoDA), ensuring that the model depends only on log-ratios among components rather than on their absolute scale.

Equivalently, the model can be rewritten as a ``log-ratio model'', i.e., a generalized linear model involving all possible pairwise log-ratios \citep{BATES2019}:

\begin{center}
\begin{equation}
g(E(Y)) = \theta_0 + \sum_{1\leq i<j\leq m} \theta_{ij} \log\left(\frac{X_i}{X_j}\right).
\label{MODELPAIRWISE}
\end{equation}
\end{center}

This representation is particularly convenient in our context, as it explicitly encodes contrasts between OTUs.

Variable selection is then carried out by fitting a penalized regression model to~\eqref{MODELPAIRWISE}. The regression coefficients $\theta_{ij}$ are estimated by minimizing a loss function $L(\theta)$ with an elastic-net penalty:

\begin{equation}
\tilde{\theta} = \arg\min_\theta \left\{ L(\theta) + \lambda_1\|\theta\|_2^2 + \lambda_2 \|\theta\|_1 \right\}.
\label{OPTPROBLEM}
\end{equation}

The penalty parameters $\lambda_1$ and $\lambda_2$ can be reparameterized in terms of $(\lambda, \alpha)$ as

$$
\lambda_1 = \frac{\lambda (1-\alpha)}{2}, \quad \lambda_2 = \lambda\alpha,
$$
where $\lambda$ controls the overall strength of the penalization and $\alpha$ determines the balance between $\ell_1$- and $\ell_2$-type regularization \citep{FRIEDMAN2010}. In the case of a linear regression model, the loss function takes the form
$$
L(\theta) = \|Y - \mathbf{M} \theta\|,
$$
where $\mathbf{M}$ is the matrix of all pairwise log-ratios, of dimension $n \times \frac{m(m-1)}{2}$. Extensions to logistic regression and other generalized linear models are discussed in \cite{FRIEDMAN2010} and apply directly in our setting.

This CoDA-based regression framework provides a global, multivariate signature of the microbiome. In the next subsection, we complement it with a pairwise log-ratio screening strategy that yields an interpretable separability index and allows for a fine-grained assessment of the discriminative power of individual log-ratios.

Step~2 in Figure~\ref{fig:overview} formalizes the compositional structure 
through a log-contrast regression model under a zero-sum constraint.
\subsection{Pairwise log-ratio screening and separability index}
\label{subsec:new_analysis}

After preprocessing and model specification, the core idea of our proposed analysis is to quantify the discriminative power of each pairwise log-ratio between OTUs. For each pair $(j, j')$ with $1 \leq j < j' \leq m$, we proceed as follows:

\begin{enumerate}
    \item \textbf{Construction of pairwise log-ratios.}  
    For each sample, we compute the pairwise log-ratio
    $$
    Z_{ijj'} = \log\left(\frac{x_{ij}}{x_{ij'}}\right).
    $$

    \item \textbf{Fitting of (multi)class regression models.}  
    We fit a logistic or multinomial logistic regression model for the response $Y$ (the group label) using $Z_{ijj'}$ as predictor, optionally adjusted for additional covariates $\mathbf{X}$.

    In the multiclass case, for $Y \in \{1, \dots, C\}$, the model takes the form, for each class $c \in \{1, \dots, C-1\}$,
    $$
    P(Y_i = c) = \frac{\exp(\beta_{0c} + \beta_{1c} Z_{ijj'} + \mathbf{X}_i^\top \boldsymbol{\gamma}_c)}{1 + \sum_{k=1}^{C-1} \exp(\beta_{0k} + \beta_{1k} Z_{ijj'} + \mathbf{X}_i^\top \boldsymbol{\gamma}_k)},
    $$
    with baseline category $Y_i = C$ given by
    $$
    P(Y_i = C) = \frac{1}{1 + \sum_{k=1}^{C-1} \exp(\beta_{0k} + \beta_{1k} Z_{ijj'} + \mathbf{X}_i^\top \boldsymbol{\gamma}_k)}.
    $$
    Parameter estimation is performed via maximum likelihood. In practice, this model can be fitted in \textsf{R} using the \texttt{multinom()} function from the \texttt{nnet} package \citep{venables2013modern}, which implements an efficient iteratively reweighted least squares (IRLS) algorithm.

    \item \textbf{Computation of pairwise AUCs.}  
    Using the fitted model, we obtain predicted probabilities for each class and evaluate the Area Under the Curve (AUC) as a measure of separability:
    \begin{itemize}
        \item For binary classification ($C=2$), we use the standard AUC,
        $$
        AUC_{jj'} = P(\text{score}^+ > \text{score}^-).
        $$
        \item For multiclass classification ($C>2$), we employ the generalized AUC proposed by \cite{hand2001simple},
        $$
        AUC_{jj'} = \frac{2}{C(C-1)} \sum_{c < c'} AUC(c, c'),
        $$
        where $AUC(c,c')$ denotes the binary AUC comparing classes $c$ and $c'$.
    \end{itemize}
\end{enumerate}

This procedure yields a symmetric $m \times m$ matrix $\mathbf{A} = [AUC_{jj'}]$, which summarizes the pairwise discriminative ability of OTU log-ratios. OTUs are then ranked according to the sum of each column of $\mathbf{A}$, prioritizing variables that consistently participate in high-scoring pairs.

\paragraph{Separability index.} 
To determine an optimal set of OTUs, we define a separability index as the average pairwise AUC among the top $k$ ranked OTUs:
$$
S_k = \frac{2}{k(k-1)} \sum_{1 \leq j < j' \leq k} AUC_{jj'}.
$$
The final relevant set is chosen up to the value $k$ at which $S_k$ reaches its maximum, providing a data-driven criterion for the number of OTUs to retain.

Steps~3 and~4 (Figure~\ref{fig:overview}) describe the pairwise log-ratio 
screening and ranking procedure based on the separability index $S_k$.

\subsection{Uncertainty quantification}

An important aspect of the proposed methodology is the estimation of the uncertainty associated with the pairwise AUCs and with the derived separability index $S_k$. Since the construction of $S_k$ involves multiple levels of aggregation, we distinguish two main sources of dependence.

\begin{enumerate}
    \item \textbf{Pairwise AUC variance in binary and multiclass settings.}  
    For each pair of OTUs $(j,j')$, a pairwise log-ratio predictor is constructed and a classification model is fitted to predict $Y$. The resulting AUC, $AUC_{jj'}$, quantifies the ability of this log-ratio to discriminate between classes.
      \item \textbf{Binary case ($C = 2$).}  
Two common approaches can be used to estimate  $\operatorname{Var}(\widehat{AUC}_{jj'})$.

      First, the classical approximation of \cite{hanley1982meaning}:
 \begin{align}
\label{eq:AUCvar_hanley}
\operatorname{Var}_{\text{Hanley}}\!\bigl(\widehat{\mathrm{AUC}}_{jj'}\bigr)
&=
\frac{
  AUC_{jj'} (1 - AUC_{jj'}) 
  + (n_j - 1)\bigl(Q_1 - AUC_{jj'}^2\bigr) 
  + (n_{j'} - 1)\bigl(Q_2 - AUC_{jj'}^2\bigr)
}{n_1 n_0},\\
Q_1 &= \frac{AUC_{jj'}}{2 - AUC_{jj'}},\; 
Q_2 = \frac{2\,AUC_{jj'}^2}{1 + AUC_{jj'}},
\end{align}
where $n_{j}$ and $n_{j'}$ denote, respectively, the number of samples from class~$j$ and class~$j'$ among the observations under comparison, that is, the samples being separated by the OTUs in $S$.

      Second, the nonparametric U-statistic approach of \cite{delong1988comparing}, for which a fast implementation is given by \cite{sun2014fast}. Define
      \[
      V_i = \frac{1}{n_0}\sum_{j=1}^{n_0} \mathbf{1}\{S_i^+ > S_j^-\},
      \quad
      W_j = \frac{1}{n_1}\sum_{i=1}^{n_1} \mathbf{1}\{S_i^+ > S_j^-\},
      \]
      where $S_i^+$ are the classifier scores on the $n_1$ positive cases and $S_j^-$ on the $n_0$ negative cases. Then
      \begin{equation}
      \label{AUCvar_delong}
      \operatorname{Var}_{\text{DeLong}}(\widehat{AUC}_{jj'}) =
      \frac{1}{n_1 (n_1 - 1)} \sum_{i=1}^{n_1} (V_i - \widehat{AUC}_{jj'})^2
      +
      \frac{1}{n_0 (n_0 - 1)} \sum_{j=1}^{n_0} (W_j -
      \widehat{AUC}_{jj'})^2.
      \end{equation}

      \item \textbf{Multiclass case ($C > 2$).}  
      When the response is multiclass, we use the Hand--Till generalization
      $$
      AUC_{jj'} = \frac{2}{C(C-1)} \sum_{c < c'} AUC_{cc'},
      $$
      where $AUC_{cc'}$ is the binary AUC comparing classes $c$ and $c'$.  The variance of $AUC_{jj'}$ is then obtained by propagating the uncertainty of its binary components:
\item 
      $$
      \operatorname{Var}(AUC_{jj'}) =
      \frac{4}{[C(C-1)]^2}
      \Bigg[
      \sum_{c < c'} \operatorname{Var}(AUC_{cc'}) +
      2 \sum_{\substack{(c < c') < (l < l') \\ (c,c') \neq (l,l')}} 
      \operatorname{Cov}(AUC_{cc'}, AUC_{ll'})
      \Bigg].
      $$
     Here, $\operatorname{Var}(AUC_{cc'})$ can be approximated using the binary variance formula in \eqref{eq:AUCvar_hanley}, whereas the covariances between binary pairs must be approximated or estimated; see below.

\item \textbf{Dependence between AUCs across OTU pairs}
Each $AUC\_{jj'}$  shares OTUs with other pairs, inducing dependence in the matrix $\mathbf{A}$. When aggregating these AUCs to compute the separability index $S_k$, this dependence propagates to its variance. 

    Recall that
    $$
    S_k = \frac{2}{k(k-1)} \sum_{j<j'\leq k} AUC_{jj'},
    $$
    so that
    $$
    \operatorname{Var}(S_k) = \frac{4}{[k(k-1)]^2} 
    \Bigg[
    \sum_{j<j'} \operatorname{Var}(AUC_{jj'}) +
    2 \sum_{\substack{(j<j') < (l<l')}} \operatorname{Cov}(AUC_{jj'}, AUC_{ll'})
    \Bigg].
    $$
    Pairs that share at least one OTU (e.g., $(AUC_{12}, AUC_{13})$) are typically positively correlated. In practice, these covariances are non-negligible but difficult to derive analytically. A feasible approximation is
    The separability index \(S_k\) is a linear combination of the pairwise AUCs within the set \(S_k\), and its variance depends not only on the variances of individual AUCs, but also on their covariances. Exact expressions would require estimating a large number of covariance terms, which is computationally burdensome and unstable.

To address this, we adopt a parsimonious approximation based on a single correlation parameter \(\rho_{\text{OTU}}\) that captures the average dependence between AUCs sharing at least one common OTU. More precisely, we write
$$
\text{Cov}\bigl(\widehat{\text{AUC}}_{jj'}, \widehat{\text{AUC}}_{kk'}\bigr)
\approx
\begin{cases}
\rho_{\text{OTU}} \, \sqrt{\widehat{\text{Var}}(\widehat{\text{AUC}}_{jj'}) \; \widehat{\text{Var}}(\widehat{\text{AUC}}_{kk'})},
& \text{if } \{j,j'\} \cap \{k,k'\} \neq \emptyset, \\
0, & \text{otherwise}.
\end{cases}
$$
Substituting this approximation into the expression for \(\text{Var}(S_k)\) yields

$$
\widehat{\text{Var}}(S_k) = \mathbf{w}_k^\top \widehat{\Sigma}_k \mathbf{w}_k,
$$

where \(\mathbf{w}_k\) is the vector of weights (all equal to \(1/[k(k-1)]\) in our case) and \(\widehat{\Sigma}_k\) is the approximate covariance matrix of the AUCs associated with pairs in \(S_k\).

In practice, \(\rho_{\text{OTU}}\) can be:
\begin{itemize}
\item Fixed to a small positive value (e.g. 0.1–0.3) based on prior experience, or  
\item Estimated from the data via a resampling scheme that compares empirical covariances of AUCs to the product of their standard errors.
\end{itemize}

In our implementation we provide both options and allow the user to perform sensitivity analyses with respect to the choice of \(\rho_{\text{OTU}}\).
    For the multiclass case, an additional layer of dependence arises because the Hand--Till AUC is itself an average of dependent binary pairs. An analogous approximation can be used for $\operatorname{Cov}(AUC_{cc'}, AUC_{ll'})$ when the class pairs share at least one class.

 \subsubsection{Alternative: non-parametric bootstrap.}  
    As an alternative to analytic variance formulas, we consider a non-parametric bootstrap procedure. This approach is particularly useful in capturing the sampling variability of $S_k$ under complex dependency structures:
    \begin{itemize}
        \item Resample the $n$ observations with replacement to generate bootstrap samples (optionally in a stratified manner by class).
        \item For each bootstrap replicate, recompute the full pairwise AUC matrix and the corresponding $S_k$.
        \item Estimate $\operatorname{Var}(S_k)$ using the empirical variance of the bootstrap replicates.
    \end{itemize}
    In the presence of class imbalance, a stratified bootstrap---resampling independently within each class---helps preserve marginal class frequencies and avoid inflated variance estimates.

Step~5 (Figure~\ref{fig:overview}) addresses variance estimation and 
uncertainty quantification of both $AUC_{jj'}$ and $S_k$.
\end{enumerate}

An additional advantage of the proposed strategy is that it naturally accommodates covariates in the multinomial regression model used to estimate class probabilities for each log-ratio pair. This allows one to adjust for potential confounding variables (such as age, sex, or clinical covariates) when assessing the discriminative ability of the OTU pairs. In the implementation provided by the \texttt{codabiocom} package, the \texttt{rowlogratios()} function accepts an optional covariate matrix $\mathbf{X}$, which is included alongside each pairwise log-ratio in the multinomial model. As a result, the computed AUC values reflect the marginal contribution of each log-ratio pair beyond the effect of included covariates, within a workflow that is fully compatible with parallel computation.

To summarize the proposed methodology and clarify its main components, Figure~\ref{fig:overview} presents an overview of the complete classification 
framework for microbiome data. Each step corresponds to one of the methodological 
sections described in detail below, from preprocessing and log-ratio modeling to 
uncertainty quantification and computational implementation.
\newpage
\begin{figure}
    \centering
\begin{tcolorbox}[
    enhanced,
    colback=blue!2!white,
    colframe=blue!60!black,
    fonttitle=\bfseries,
    title={Overview of the Proposed Classification Framework for Microbiome Data},
    boxrule=0.8pt,
    arc=3mm,
    breakable,
    before skip=5pt,
    after skip=5pt
]
\begin{center}
\textbf{Objective:} Identify differential OTUs that best discriminate between groups $(g = 1,\ldots,k)$ 
using compositional data analysis and pairwise log-ratio modeling.
\end{center}

\begin{tikzpicture}[
    node distance=0.3cm,
    every node/.style={align=justify, font=\small, text width=15cm, rounded corners, inner sep=3pt}
]

\node (start) [draw, fill=gray!10]
    {\textbf{Input data:} Abundance matrix $\mathbf{X}_{n\times m}$ where rows = samples, columns = OTUs, and class labels $Y \in \{1,\ldots,k\}$.};

\node (preproc) [draw, fill=green!10, below=of start]
    {\textbf{1. Data Preprocessing:}\\
    (a) Remove OTUs with $\sum_i x_{ij}\leq 2$;\\
    (b) Replace zeros using Bayesian-multiplicative imputation \citep{BAYZEROSIMP};\\
    (c) Apply centered log-ratio (clr) transformation.};

\node (model) [draw, fill=orange!10, below=of preproc]
    {\textbf{2. Log-Contrast Regression Model:}\\
    Model the categorical response $Y$ via a penalized log-contrast regression:
    \[
    g(E[Y]) = \alpha_0 + \sum_{i=1}^m \alpha_i \log(X_i), 
    \quad \text{s.t. } \sum_i \alpha_i=0.
    \]
    For $k>2$, use multinomial logistic link $g(\cdot)$.
    \textit{Equivalent form:} pairwise log-ratio model
    \[
    g(E[Y]) = \theta_0 + \sum_{i<j} \theta_{ij} \log\!\left(\frac{X_i}{X_j}\right),
    \]
    estimated using elastic-net penalization \citep{FRIEDMAN2010}.};

\node (new) [draw, fill=cyan!10, below=of model]
    {\textbf{3. Pairwise Log-Ratio Screening:}\\
    For each OTU pair $(j,j')$:
    \begin{enumerate}
        \item Compute log-ratio $Z_{ijj'}=\log(x_{ij}/x_{ij'})$;
        \item Fit logistic or multinomial model $Y \sim Z_{ijj'} + \mathbf{X}$;
        \item Compute the AUC ($AUC_{jj'}$) for class discrimination.
    \end{enumerate}
    Obtain the symmetric AUC matrix $\mathbf{A}=[AUC_{jj'}]$.};

\node (ranking) [draw, fill=purple!10, below=of new]
    {\textbf{4. OTU Ranking and Selection:}\\
    Rank OTUs by total AUC contribution. Define separability index:
    \[
    S_k = \frac{2}{k(k-1)} \sum_{1\leq j<j'\leq k} AUC_{jj'},
    \]
    and select the number $k$ that maximizes $S_k$.};

\node (var) [draw, fill=red!10, below=of ranking]
    {\textbf{5. Variance and Uncertainty Estimation:}\\
    Estimate $\operatorname{Var}(AUC_{jj'})$ using:
    \begin{itemize}
        \item Hanley–McNeil (\cite{hanley1982meaning}) or DeLong (\cite{delong1988comparing}) formulas;
        \item Propagation to $\operatorname{Var}(S_k)$ including OTU-level covariance;
        \item Alternatively, stratified bootstrap resampling.
    \end{itemize}};

\node (end) [draw, fill=blue!10, below=of var]
    {\textbf{6. Output:}\\
    $\Rightarrow$ Optimal subset of differential OTUs maximizing separability $S_k$;\\
    $\Rightarrow$ Statistical uncertainty quantified by analytical or bootstrap variance.};
\end{tikzpicture}
\centering
\textbf{Software:} Implemented in the \texttt{codabiocom} R package, supporting covariate adjustment and parallel processing.
\end{tcolorbox}
\caption{Overview of the proposed classification framework for microbiome data.}
\label{fig:overview}
\end{figure}
\clearpage

As outlined in Figure~\ref{fig:overview}, the workflow proceeds in six main steps: 
data preprocessing, log-contrast modeling, pairwise log-ratio screening, 
ranking and selection via the separability index $S_k$, uncertainty quantification, 
and final output generation. 

\subsection{Computational complexity and parallel implementation}
\label{subsec:complexity}

The pairwise log-ratio approach inherently involves a computational cost that grows quadratically with the number of OTUs, since the total number of unique pairs is $\binom{m}{2} = O(m^2)$. This can become a bottleneck when working with large, high-dimensional microbiome datasets.

To address this, the pairwise AUC computation has been implemented in a fully parallelized framework using the \texttt{foreach} and \texttt{doParallel} packages in \textsf{R}. The core operation is handled by the \texttt{rowlogratios()} function, which fits the (multinomial) regression model and computes the pairwise AUC for each log-ratio. The \texttt{calcAUClr()} function orchestrates the parallel computation across all OTU pairs, efficiently distributing tasks across multiple cores and thus exploiting available computational resources.

The final step (Step~6 in Figure~\ref{fig:overview}) focuses on computational 
scalability and parallelization, ensuring the feasibility of the method for 
high-dimensional microbiome datasets.

This design enables not only the calculation of the initial AUC matrix, but also makes bootstrap resampling of the separability index $S_k$ feasible at scale. By resampling the data and recomputing the entire AUC matrix in each replicate, this strategy naturally captures all levels of dependence: among OTU pairs, within multiclass combinations, and across the final index calculation.

All steps are implemented in the open-source \texttt{codabiocom} \textsf{R} package, which is freely available at
\begin{center}
\url{https://github.com/Cruzalirio/codabiocom}
\end{center}
ensuring that the proposed methodology is fully reproducible and scalable to large compositional datasets.



\section{Application to microbiome data}
\label{sec:application}

\subsection{Data description}

We illustrate the proposed methodology using the colorectal cancer screening dataset originally reported by \cite{baxter2016microbiota} and later reanalyzed by \cite{greenacre2021compositional}. The data consist of operational taxonomic unit (OTU) abundances for 336 microbial features measured on stool samples from subjects classified as adenoma or control. In addition to the compositional microbiome data, the dataset includes several covariates, such as age, gender, and diabetes medication usage.

Our primary objective is to classify subjects into adenoma versus control using the compositional OTU data, while appropriately adjusting for potential confounders. To this end, we applied the \texttt{LRRelev} procedure, which combines log-ratio logistic regression with the Hanley separation index. Four different modeling strategies were considered:

\begin{itemize}
    \item \textbf{Model A}: unadjusted (no covariates),
    \item \textbf{Model B}: adjusted for age,
    \item \textbf{Model C}: adjusted for age and gender,
    \item \textbf{Model D}: adjusted for age, gender, and diabetes medication.
\end{itemize}

In all cases, the OTU data were transformed into pairwise log-ratios using the \texttt{rowlogratios()} function in \texttt{codabiocom}, and the corresponding separation indices were computed following the methodology described in the previous sections. All computations were performed using six processing cores to ensure adequate computational efficiency.

\subsection{OTU relevance across models}

Each model produces a relevance ranking of OTUs based on the separability index, with confidence intervals derived from the estimated variance. Table~\ref{tab:otus-relevantes} summarizes, for each model, the number of relevant OTUs selected and the top ranked taxa.

\begin{table}[H]
\centering
\caption{Number and top relevant OTUs (ordered by association index) per model.}
\label{tab:otus-relevantes}
\begin{tabular}{|c|c|p{9cm}|}
\hline
\textbf{Model} & \textbf{Relevant OTUs} & \textbf{Top OTUs (ordered)} \\
\hline
A (Unadjusted) & 7 & Otu000105, Otu000310, Otu000281, Otu000264, Otu000058, Otu000113, Otu000067 \\
B (Age only) & 17 & Otu000310, Otu000105, Otu000281, Otu000264, Otu000113, Otu000260, Otu000286, Otu000058, Otu000067, Otu000057, Otu000356, Otu000097, Otu000094, Otu000053, Otu000297, Otu000054, Otu000278 \\
C (Age, Gender) & 9 & Otu000310, Otu000105, Otu000281, Otu000264, Otu000113, Otu000260, Otu000286, Otu000058, Otu000067 \\
D (Age, Gender, Diabetes) & 9 & Otu000310, Otu000105, Otu000281, Otu000264, Otu000113, Otu000260, Otu000286, Otu000058, Otu000067 \\
\hline
\end{tabular}
\end{table}

The consistency in the selection of Otu000310, Otu000105, Otu000281, Otu000264, Otu000113, Otu000067, and Otu000058 across all models highlights their robustness as discriminant features for diagnosis classification. The inclusion of covariates such as age, gender, and diabetes medication leads to changes in the number of selected OTUs and affects their ranking, suggesting that some taxa may be partially confounded by demographic or clinical factors.

Table~\ref{tab:sep-index-ci} summarizes the estimated separation index and its confidence interval for each model.

\begin{table}[H]
\centering
\small
\caption{Separation index and 95\% confidence intervals per model.}
\label{tab:sep-index-ci}
\begin{tabular}{lrrrr}
\toprule
Model & Relevant OTUs & Separation index & Lower CI & Upper CI \\
\midrule
A (Unadjusted) & 7  & 0.5816 & 0.567 & 0.596 \\
B (Age only)   & 17 & 0.6685 & 0.655 & 0.682 \\
C (Age, Gender) & 9 & 0.6962 & 0.683 & 0.709 \\
D (Age, Gender, Diabetes) & 9 & 0.7001 & 0.687 & 0.713 \\
\bottomrule
\end{tabular}
\end{table}

The separation index estimates and their associated confidence intervals provide insight into the robustness and precision of OTU-based discrimination between adenoma and control groups. Model~A, which does not adjust for any covariates, shows moderate separation performance, with a separation index around $0.58$. Adjustment for demographic covariates substantially improves discrimination: Model~B, adjusting only for age, increases the separation index to approximately $0.67$, while Models~C and~D (adjusting for age and gender, and for age, gender, and diabetes medication, respectively) yield separation indices close to 0.70,  representing the strongest overall discrimination among the four models.

\section{Discussion}

The findings of this study highlight several important aspects of the proposed framework for identifying discriminant OTUs in compositional microbiome data.  

First, a core set of taxa consistently emerged across all models, even after adjusting for demographic and clinical covariates. This consistency indicates that these microbial features are robustly associated with disease status and are likely to represent biologically meaningful signals rather than artifacts of data preprocessing or model specification.

Second, the influence of covariates on model performance and feature selection is evident. Accounting for variables such as age, gender, and diabetes medication improves model precision—reflected in narrower confidence intervals for the separability index $S_k$—and reveals additional taxa that were not detected in unadjusted analyses. This result reinforces the importance of incorporating relevant covariates in microbiome studies to reduce confounding and to improve the interpretability of microbial associations with health outcomes.

Third, the proposed framework adds an element of statistical interpretability that complements existing ordination-based techniques such as Correspondence Analysis (CA) and Canonical Correspondence Analysis (CCA).  
The Hanley separation index provides an intuitive, non-parametric measure of group separation that directly quantifies discriminatory power, while the associated confidence intervals enable formal inference. This combination bridges exploratory visualization and rigorous statistical modeling, making the results both interpretable and statistically grounded.

From a biological standpoint, the persistence of certain taxa across multiple adjusted models suggests that these organisms may play a genuine role in disease development or progression. Their stability across analytical settings supports their potential use as biomarkers in diagnostic or preventive microbiome research, and as targets for future mechanistic or interventional studies.

Taken together, these results demonstrate that integrating log-ratio modeling with covariate adjustment and uncertainty quantification provides a statistically principled and biologically meaningful approach for OTU selection in compositional data. The proposed methodology enhances both interpretability and reproducibility, which are essential for advancing microbiome-based inference and discovery.

\section{Conclusions and future work}

This study presents a unified framework for multinomial classification and feature selection in compositional microbiome data.  
By combining log-ratio transformations, penalized regression, and pairwise AUC-based separability measures, the approach offers a balance between computational feasibility, interpretability, and statistical rigor. The separability index $S_k$ provides a simple yet effective way to rank and select discriminant taxa, while analytic and bootstrap-based uncertainty estimates ensure that the conclusions drawn are statistically robust.

Our application to the Baxter dataset confirms that the proposed approach not only reproduces known microbial patterns but also enhances inference by adjusting for covariates and providing confidence intervals. In this sense, the method complements ordination-based tools by offering a probabilistic perspective that directly connects microbial compositions to clinical outcomes.

Looking ahead, several extensions are worth pursuing. One direction involves extending the framework to hierarchical taxonomic levels (e.g., genus or family), which could improve biological interpretation. Another promising line is the incorporation of Bayesian priors to stabilize feature selection in ultra–high-dimensional contexts. Scalability to other omics domains—such as metagenomics, metabolomics, or transcriptomics—will also broaden the applicability of the method. Finally, integrating cross-study validation procedures could help assess the generalizability and reproducibility of the identified microbial signatures.

Overall, the proposed framework provides a transparent, reproducible, and statistically grounded approach for compositional data classification, implemented in the open-source \texttt{codabiocom} \textsf{R} package. It represents a step toward more interpretable and reproducible microbiome analytics, bridging methodological rigor with biological insight.

\newcommand{\etalchar}[1]{$^{#1}$}

\end{document}